\documentclass[12pt]{article}
\usepackage{graphicx}
\usepackage{natbib}
\usepackage{graphicx}   
\usepackage{xspace}     
\usepackage{amsmath}
\usepackage{hyperref}
\usepackage{color}
\usepackage{multirow}
\usepackage{array}
\usepackage{amssymb}
\RequirePackage{lineno}
\newcolumntype{C}[1]{>{\centering\let\newline\\\arraybackslash\hspace{0pt}}m{#1}}

\newcommand{\pT}{$p_T$\xspace}

\newcommand{\RAA}{$R_{AA}$\xspace}

\newcommand{\pp}{$p$+$p$\xspace}

\setcitestyle{numbers}

\newcommand\pubnumber{CIPANP2018-Nattrass}
\newcommand\pubdate{\today}

\def\napoli{Department of Physics and Astronomy\\
University of Tennessee, Knoxville, TN, USA-37996 USA}

\def\Title#1{\begin{center} {\Large #1 } \end{center}}
\def\Author#1{\begin{center}{ \sc #1} \end{center}}
\def\Address#1{\begin{center}{ \it #1} \end{center}}

\newcommand\pubblock{\rightline{\begin{tabular}{l} \pubnumber\\
         \pubdate  \end{tabular}}}
\newenvironment{Abstract}{\begin{quotation}  }{\end{quotation}}
\newenvironment{Presented}{\begin{quotation} \begin{center} 
             PRESENTED AT\end{center}\bigskip 
      \begin{center}\begin{large}}{\end{large}\end{center} \end{quotation}}
\def\Acknowledgements{\bigskip  \bigskip \begin{center} \begin{large}
             \bf ACKNOWLEDGEMENTS \end{large}\end{center}}

\def\beq{\begin{equation}}
\def\eeq#1{\label{#1}\end{equation}}
\def\eeqn{\end{equation}}

\def\beqa{\begin{eqnarray}}
\def\eeqa#1{\label{#1}\end{eqnarray}}
\def\eeqan{\end{eqnarray}}

\let\bar=\overbar

\def\Dslash{\not{\hbox{\kern-4pt $D$}}}
\def\dslash{\not{\hbox{\kern-2pt $\del$}}}

\def\msb{{\bar{\ssstyle M \kern -1pt S}}}

\begin{document}
\begin{titlepage}
\pubblock

\vfill
\Title{Measurements of jets in heavy ion collisions}
\vfill
\Author{Christine Nattrass}
\Address{\napoli}
\vfill
\begin{Abstract}
The Quark-Gluon Plasma (QGP) is created in high energy heavy ion collisions at the Relativistic Heavy Ion Collider (RHIC) and the Large Hadron Collider (LHC).  This medium is transparent to electromagnetic probes but nearly opaque to colored probes.  Hard partons produced early in the collision fragment and hadronize into a collimated spray of particles called a jet.  The partons lose energy as they traverse the medium, a process called jet quenching.  Most of the lost energy is still correlated with the parent parton, contributing to particle production at larger angles and lower momenta relative to the parent parton than in proton-proton collisions.  This partonic energy loss can be measured through several observables, each of which give different insights into the degree and mechanism of energy loss.  The measurements to date are summarized and the path forward is discussed.
\end{Abstract}
\vfill
\begin{Presented}
Thirteenth Conference on the Intersections of Particle and Nuclear Physics\\
Indian Wells, CA, May 28--June 3, 2018
\end{Presented}
\vfill
\end{titlepage}
\def\thefootnote{\fnsymbol{footnote}}
\setcounter{footnote}{0}

\section{Introduction}

The QGP produced in high energy nuclear collisions is transparent to electromagnetic probes but nearly opaque to colored probes.  Hard partons which traverse the medium lose a large fraction of their energy through a combination of bremsstrahlung and collisional energy loss, a process called jet quenching.  These hard partons then fragment into a collimated spray of particles called a jet.  The lost energy partially retains its correlation with the parent parton, although it is slightly further from the jet axis on average than it would have been without interactions with the medium.   
In~\cite{Connors:2017ptx} we reviewed the evidence for jet quenching in a QGP, which is in qualitative agreement with partonic energy loss and concomittant broadening of the jet and softening of the distribution of final state partons.

The question now is how to use our extensive measurements to quantitatively determine the properties of the medium.  There are recent investigations of alternate observables to see if these may be more sensitive to the properties of the medium.  However, the field also needs to investigate the impact of experimental techniques on the sensitivity of various measurements and possible biases introduced by these methods.  This process began in a research workshop on the Definition of Jets in a Large Background~\cite{BackgroundWorkshop}.  There are also extensive efforts underway by the JETSCAPE collaboration~\cite{JETSCAPE} to provide a framework which incorporates both partonic energy loss and a realistic medium, to be followed by a Bayesian analysis comparing measurements to these models to optimize our constraints of medium properties.

\section{Summary of measurements of jets}

In~\cite{Connors:2017ptx} we divided evidence for jet quenching into evidence that there is energy loss and evidence that this energy is redistributed slightly further from the jet axis.  Partonic energy loss can be observed through the measurement of the nuclear modification factor
\begin{equation}
 R_{AA} = \frac{\sigma_{NN}}{\langle N_{bin}\rangle} \frac{d^2N_{AA}/dp_{T}d\eta}{d^2\sigma_{pp}/dp_{T}d\eta}
\end{equation} where \pT is the transverse momentum, $\langle N_{bin}\rangle$ is the average number of binary nucleon-nucleon collisions for a given range of impact parameter, $\sigma_{NN}$ is the integrated nucleon-nucleon cross section, and $\eta$ is the pseudorapidity.  $N_{AA}$ and  $\sigma_{pp}$ in this context are the yield in $AA$ collision and cross section in \pp collisions for a particular observable.  The high \pT~particle cross-section should scale with the number of binary collisions and therefore \RAA~=~1 if nucleus-nucleus collisions were simply a superposition of nucleon-nucleon collisions.  An \RAA~$<$~1 indicates suppression and an \RAA~$>$~1 indicates enhancement.  The single hadron \RAA is observed to be signficantly suppressed both at at the Relativistic Heavy Ion Collider (RHIC)~\cite{Adams:2003kv,Adler:2003qi,Back:2004bq} and the Large Hadron Collider (LHC)~\cite{Aamodt:2010jd,CMS:2012aa,CMS-PAS-HIN-15-015}.  Similar results are observed for the \RAA of reconstructed jets~\cite{Adam:2015ewa,Aad:2014bxa,Khachatryan:2016jfl} and through a number of other observables~\cite{Adams:2003im,Nattrass:2018eyk,Aad:2010bu,Chatrchyan:2011sx}.

The lost energy on average retains some correlation with the jet axis, with bremsstrahlung and radiative energy loss leading to a slightly broader jet with lower momentum constituents on average.  This can be quantified by directly measuring the fragmentation function, the momentum distribution of particles in jets, which is indeed observed to be slightly softer~\cite{Aad:2014wha,Chatrchyan:2014ava,Chatrchyan:2012gw,Adare:2010yw}.  It can also be measured through other observables~\cite{Agakishiev:2011st,Khachatryan:2016erx,Adamczyk:2013jei,Adamczyk:2016fqm,Chatrchyan:1605718}, which lead to qualitatively consistent results.  Recently, a number of other observables have been explored to attempt to identify those which are most sensitive to the medium~\cite{Acharya:2018uvf,Acharya:2017goa,CMS:Splitting}.  It is not yet clear whether any of these observables are more sensitive to the properties of the medium and, if so, which are most sensitive.

\section{How to move forward}

Overall, there are extensive measurements which, in principle, could constrain theoretical models for jet quenching.  However, so far there has only been one detailed comparison between single particle \RAA and a range of models done by the JET collaboration to constrain the jet transport coefficient $\hat{q} = Q^2/L$, where $L$ is the path-length traversed and $Q$ is the transverse momentum lost to the medium~\cite{Burke:2013yra}.  Detailed comparisons between theory and experiment so far have been limited by a mismatch between what theoretical calculations can predict and what experiments can measure.  To move forward, we need to make sure that we fully understand the impact of the background and possible biases in measurements, we need to view measurements comprehensively rather than in isolation, and we need to seriously evaluate which observables are most sensitive to the properties of the medium to guide future studies.

Jets are complicated and their definition is non-trivial, even in the absence of a large background.  The Snowmass Accord was essential in order to quantitatively understand jet production in elemenary collisions and stipulated that theorists and experimentalistis must use the same jet finder~\cite{Huth:1990mi}.  The jet finder effectively defines the jet.  The experimental techniques used in heavy ion collisions which impose strong biases may therefore be understood as part of the definition of the jet.  A workshop was held at Brookhaven National Laboratory to attempt to come to an agreement on the definition of jets in a large background~\cite{BackgroundWorkshop}, to be followed by a position paper outlining points of agreement in the field.

Furthermore, experimental techniques may lead to biases in the population of jets.  Measurements in heavy ion collisions generally focus on narrow jets and the background may be suppressed by selecting jets which fragment into higher momentum hadrons.  Since quark jets are narrower and fragment harder than gluon jets~\cite{OPAL:1995ab,Abreu:1995hp}, this explicitly favors reconstruction of quark jets over gluon jets.  This may also select jets which have interacted less with the medium.  Interactions with the medium may also lead to large differences in the degree of modification with changing momentum.  Therefore each measurement needs to be interpreted as a piece of the puzzle rather than being viewed in isolation.

Finally, while the search for new observables is promising, it is still unclear which observables are best.  Many observables are sensitive to the same effects and therefore to the same medium properties.  The most useful observables for constraining the properties of the medium may not be novel but rather straightforward observables which can be measured to high precision.

Work by the JETSCAPE collaboration will be key for answering all of these questions.  JETSCAPE is developing a framework which incorporates with both a realistic medium and realistic partonic energy loss in Monte Carlo models.  It will be possible to fully implement experimental techniques in the JETSCAPE Monte Carlo models so that biases from the experimental technique can be reproduced in the theory.  The JETSCAPE Monte Carlo models will subsequently be used for a systematic comparison to multiple measurements at once using a Bayesian analysis similar to those done for bulk observables~\cite{Novak:2013bqa,Bernhard:2016tnd}, allowing quantitative constraints of the properties of the medium using measurements of jets.  This can also be used to test the impact of background subtraction methods and determine which observables are most sensitive to the properties of the medium.

\section{Conclusions}
The field has amassed a wide range of evidence for jet quenching and this evidence is in qualitative agreement with most models.  So far, however, there have been few quantitative constraints on the properties of the medium from these measurements.  Quantitative constraints require acknowledging the limitations of measurements to date, including biases and the possible impact of background.  This will be aided significantly by the efforts of the JETSCAPE collaboration.

\Acknowledgements
I am thankful to Megan Connors, Rosi Reed, and Sevil Salur for many useful discussions and to Redmer Bertens for comments on the manuscript.
This material is based upon work supported by the Division of Nuclear Physics of the U.S. Department of Energy under Grant No. DE-FG02-96ER40982.



\begin{thebibliography}{34}
\providecommand{\natexlab}[1]{#1}
\providecommand{\url}[1]{\texttt{#1}}
\expandafter\ifx\csname urlstyle\endcsname\relax
  \providecommand{\doi}[1]{doi: #1}\else
  \providecommand{\doi}{doi: \begingroup \urlstyle{rm}\Url}\fi

\bibitem[Connors et~al.(2018)Connors, Nattrass, Reed, and
  Salur]{Connors:2017ptx}
Megan Connors, Christine Nattrass, Rosi Reed, and Sevil Salur.
\newblock {Jet measurements in heavy ion physics}.
\newblock \emph{Rev. Mod. Phys.}, 90:\penalty0 025005, 2018.
\newblock \doi{10.1103/RevModPhys.90.025005}.

\bibitem[Bac(2018)]{BackgroundWorkshop}
Workshop on the definition of jets in a large background, 2018.
\newblock URL \url{https://www.bnl.gov/jets18}.

\bibitem[JET(2018)]{JETSCAPE}
Jetscape collaboration website, 2018.
\newblock URL \url{http://jetscape.wayne.edu/}.

\bibitem[Adams et~al.(2003{\natexlab{a}})]{Adams:2003kv}
J.~Adams et~al.
\newblock {Transverse momentum and collision energy dependence of high p(T)
  hadron suppression in Au+Au collisions at ultrarelativistic energies}.
\newblock \emph{Phys.Rev.Lett.}, 91:\penalty0 172302, 2003{\natexlab{a}}.
\newblock \doi{10.1103/PhysRevLett.91.172302}.

\bibitem[Adler et~al.(2003)]{Adler:2003qi}
S.S. Adler et~al.
\newblock {Suppressed $\pi^0$ production at large transverse momentum in
  central Au+ Au collisions at S(NN)**1/2 = 200 GeV}.
\newblock \emph{Phys.Rev.Lett.}, 91:\penalty0 072301, 2003.
\newblock \doi{10.1103/PhysRevLett.91.072301}.

\bibitem[Back et~al.(2004)]{Back:2004bq}
B.B. Back et~al.
\newblock {Pseudorapidity dependence of charged hadron transverse momentum
  spectra in d+Au collisions at s(NN)**(1/2) = 200 GeV}.
\newblock \emph{Phys.Rev.}, C70:\penalty0 061901, 2004.
\newblock \doi{10.1103/PhysRevC.70.061901}.

\bibitem[Aamodt et~al.(2011)]{Aamodt:2010jd}
K.~Aamodt et~al.
\newblock {Suppression of Charged Particle Production at Large Transverse
  Momentum in Central Pb--Pb Collisions at $\sqrt{s_{NN}} = 2.76$ TeV}.
\newblock \emph{Phys.Lett.}, B696:\penalty0 30--39, 2011.
\newblock \doi{10.1016/j.physletb.2010.12.020}.

\bibitem[Chatrchyan et~al.(2012{\natexlab{a}})]{CMS:2012aa}
Serguei Chatrchyan et~al.
\newblock {Study of high-pT charged particle suppression in PbPb compared to
  $pp$ collisions at $\sqrt{s_{NN}}=2.76$ TeV}.
\newblock \emph{Eur.Phys.J.}, C72:\penalty0 1945, 2012{\natexlab{a}}.
\newblock \doi{10.1140/epjc/s10052-012-1945-x}.

\bibitem[CMS(2016)]{CMS-PAS-HIN-15-015}
{Measurement of the charged particle nuclear modification factor in PbPb
  collisions at sqrt(sNN) = 5.02 TeV}, 2016.
\newblock URL \url{cds.cern.ch/record/2155301}.
\newblock CMS-PAS-HIN-15-015.

\bibitem[Adam et~al.(2015)]{Adam:2015ewa}
Jaroslav Adam et~al.
\newblock {Measurement of jet suppression in central Pb-Pb collisions at
  $\sqrt{s_{\rm NN}}$ = 2.76 TeV}.
\newblock \emph{Phys.Lett.}, B746:\penalty0 1--14, 2015.
\newblock \doi{10.1016/j.physletb.2015.04.039}.

\bibitem[Aad et~al.(2015)]{Aad:2014bxa}
Georges Aad et~al.
\newblock {Measurements of the Nuclear Modification Factor for Jets in Pb+Pb
  Collisions at $\sqrt{s_{\mathrm{NN}}}=2.76$ TeV with the ATLAS Detector}.
\newblock \emph{Phys.Rev.Lett.}, 114\penalty0 (7):\penalty0 072302, 2015.
\newblock \doi{10.1103/PhysRevLett.114.072302}.

\bibitem[Khachatryan et~al.(2017)]{Khachatryan:2016jfl}
Vardan Khachatryan et~al.
\newblock {Measurement of inclusive jet cross sections in $pp$ and PbPb
  collisions at $\sqrt{s_{NN}}=$ 2.76 TeV}.
\newblock \emph{Phys. Rev.}, C96\penalty0 (1):\penalty0 015202, 2017.
\newblock \doi{10.1103/PhysRevC.96.015202}.

\bibitem[Adams et~al.(2003{\natexlab{b}})]{Adams:2003im}
J.~Adams et~al.
\newblock {Evidence from d + Au measurements for final state suppression of
  high p(T) hadrons in Au+Au collisions at RHIC}.
\newblock \emph{Phys. Rev. Lett.}, 91:\penalty0 072304, 2003{\natexlab{b}}.
\newblock \doi{10.1103/PhysRevLett.91.072304}.

\bibitem[Nattrass(2018)]{Nattrass:2018eyk}
Christine Nattrass.
\newblock {Reexamining the iconic dihadron correlation measurement
  demonstrating jet quenching}.
\newblock \emph{Phys. Rev.}, C97\penalty0 (3):\penalty0 034916, 2018.
\newblock \doi{10.1103/PhysRevC.97.034916}.

\bibitem[Aad et~al.(2010)]{Aad:2010bu}
Georges Aad et~al.
\newblock {Observation of a Centrality-Dependent Dijet Asymmetry in Lead-Lead
  Collisions at $\sqrt{s_{NN}}=2.77$ TeV with the ATLAS Detector at the LHC}.
\newblock \emph{Phys.Rev.Lett.}, 105:\penalty0 252303, 2010.
\newblock \doi{10.1103/PhysRevLett.105.252303}.

\bibitem[Chatrchyan et~al.(2011)]{Chatrchyan:2011sx}
Serguei Chatrchyan et~al.
\newblock {Observation and studies of jet quenching in PbPb collisions at
  nucleon-nucleon center-of-mass energy = 2.76 TeV}.
\newblock \emph{Phys. Rev.}, C84:\penalty0 024906, 2011.
\newblock \doi{10.1103/PhysRevC.84.024906}.

\bibitem[Aad et~al.(2014)]{Aad:2014wha}
Georges Aad et~al.
\newblock {Measurement of inclusive jet charged-particle fragmentation
  functions in Pb+Pb collisions at $\sqrt{s_{NN}} = 2.76$ TeV with the ATLAS
  detector}.
\newblock \emph{Phys.Lett.}, B739:\penalty0 320--342, 2014.
\newblock \doi{10.1016/j.physletb.2014.10.065}.

\bibitem[Chatrchyan et~al.(2014)]{Chatrchyan:2014ava}
Serguei Chatrchyan et~al.
\newblock {Measurement of jet fragmentation in PbPb and pp collisions at
  $\sqrt{s_{NN}}=2.76$ TeV}.
\newblock \emph{Phys.Rev.}, C90\penalty0 (2):\penalty0 024908, 2014.
\newblock \doi{10.1103/PhysRevC.90.024908}.

\bibitem[Chatrchyan et~al.(2012{\natexlab{b}})]{Chatrchyan:2012gw}
Serguei Chatrchyan et~al.
\newblock {Measurement of jet fragmentation into charged particles in $pp$ and
  PbPb collisions at $\sqrt{s_{NN}}=2.76$ TeV}.
\newblock \emph{JHEP}, 1210:\penalty0 087, 2012{\natexlab{b}}.
\newblock \doi{10.1007/JHEP10(2012)087}.

\bibitem[Adare et~al.(2010)]{Adare:2010yw}
A.~Adare et~al.
\newblock {High $p_T$ direct photon and $\pi^0$ triggered azimuthal jet
  correlations and measurement of $k_T$ for isolated direct photons in $p+p$
  collisions at $sqrt{s}=200$ GeV}.
\newblock \emph{Phys.Rev.}, D82:\penalty0 072001, 2010.
\newblock \doi{10.1103/PhysRevD.82.072001}.

\bibitem[Agakishiev et~al.(2012)]{Agakishiev:2011st}
G.~Agakishiev et~al.
\newblock {System size and energy dependence of near-side di-hadron
  correlations}.
\newblock \emph{Phys.Rev.}, C85:\penalty0 014903, 2012.
\newblock \doi{10.1103/PhysRevC.85.014903}.

\bibitem[Khachatryan et~al.(2016)]{Khachatryan:2016erx}
Vardan Khachatryan et~al.
\newblock {Correlations between jets and charged particles in PbPb and pp
  collisions at $ \sqrt{s_{\mathrm{NN}}}=2.76 $ TeV}.
\newblock \emph{JHEP}, 02:\penalty0 156, 2016.
\newblock \doi{10.1007/JHEP02(2016)156}.

\bibitem[Adamczyk et~al.(2014)]{Adamczyk:2013jei}
L.~Adamczyk et~al.
\newblock {Jet-Hadron Correlations in $\sqrt{s_{NN}} = 200$ GeV $p+p$ and
  Central $Au+Au$ Collisions}.
\newblock \emph{Phys.Rev.Lett.}, 112\penalty0 (12):\penalty0 122301, 2014.
\newblock \doi{10.1103/PhysRevLett.112.122301}.

\bibitem[Adamczyk et~al.(2017)]{Adamczyk:2016fqm}
L.~Adamczyk et~al.
\newblock {Dijet imbalance measurements in $Au+Au$ and $pp$ collisions at
  $\sqrt{s_{NN}} = 200$  GeV at STAR}.
\newblock \emph{Phys. Rev. Lett.}, 119\penalty0 (6):\penalty0 062301, 2017.
\newblock \doi{10.1103/PhysRevLett.119.062301}.

\bibitem[Chatrchyan et~al.(2013)]{Chatrchyan:1605718}
Serguei Chatrchyan et~al.
\newblock {Modification of jet shapes in PbPb collisions at $\sqrt{s_{NN}}$ =
  2.76 TeV}.
\newblock \emph{Phys. Lett. B}, 730\penalty0 (arXiv:1310.0878. CMS-HIN-12-002.
  CERN-PH-EP-2013-189):\penalty0 243. 31 p, Oct 2013.

\bibitem[Acharya et~al.(2018{\natexlab{a}})]{Acharya:2018uvf}
Shreyasi Acharya et~al.
\newblock {Medium modification of the shape of small-radius jets in central
  Pb-Pb collisions at $\sqrt{s_{\mathrm {NN}}} = 2.76\,\rm{TeV}$}.
\newblock \emph{Submitted to: JHEP}, 2018{\natexlab{a}}.

\bibitem[Acharya et~al.(2018{\natexlab{b}})]{Acharya:2017goa}
S.~Acharya et~al.
\newblock {First measurement of jet mass in Pb–Pb and p–Pb collisions at
  the LHC}.
\newblock \emph{Phys. Lett.}, B776:\penalty0 249--264, 2018{\natexlab{b}}.
\newblock \doi{10.1016/j.physletb.2017.11.044}.

\bibitem[Sirunyan et~al.(2017)]{CMS:Splitting}
Albert~M Sirunyan et~al.
\newblock {Measurement of the splitting function in pp and PbPb collisions at
  $\sqrt{s_{_{\mathrm{NN}}}}=$ 5.02 TeV}.
\newblock 2017.

\bibitem[Burke et~al.(2014)]{Burke:2013yra}
Karen~M. Burke et~al.
\newblock {Extracting the jet transport coefficient from jet quenching in
  high-energy heavy-ion collisions}.
\newblock \emph{Phys.Rev.}, C90\penalty0 (1):\penalty0 014909, 2014.
\newblock \doi{10.1103/PhysRevC.90.014909}.

\bibitem[Huth et~al.(1990)]{Huth:1990mi}
John~E. Huth et~al.
\newblock {Toward a standardization of jet definitions}.
\newblock In \emph{{1990 DPF Summer Study on High-energy Physics: Research
  Directions for the Decade (Snowmass 90) Snowmass, Colorado, June 25-July 13,
  1990}}, pages 0134--136, 1990.
\newblock URL \url{http://lss.fnal.gov/cgi-bin/find_paper.pl?conf-90-249}.

\bibitem[Akers et~al.(1995)]{OPAL:1995ab}
R.~Akers et~al.
\newblock {A Model independent measurement of quark and gluon jet properties
  and differences}.
\newblock \emph{Z. Phys.}, C68:\penalty0 179--202, 1995.

\bibitem[Abreu et~al.(1996)]{Abreu:1995hp}
P.~Abreu et~al.
\newblock {Energy dependence of the differences between the quark and gluon jet
  fragmentation}.
\newblock \emph{Z. Phys.}, C70:\penalty0 179--196, 1996.
\newblock \doi{10.1007/s002880050095}.

\bibitem[Novak et~al.(2014)Novak, Novak, Pratt, Vredevoogd, Coleman-Smith, and
  Wolpert]{Novak:2013bqa}
John Novak, Kevin Novak, Scott Pratt, Joshua Vredevoogd, Chris Coleman-Smith,
  and Robert Wolpert.
\newblock {Determining Fundamental Properties of Matter Created in
  Ultrarelativistic Heavy-Ion Collisions}.
\newblock \emph{Phys. Rev.}, C89\penalty0 (3):\penalty0 034917, 2014.
\newblock \doi{10.1103/PhysRevC.89.034917}.

\bibitem[Bernhard et~al.(2016)Bernhard, Moreland, Bass, Liu, and
  Heinz]{Bernhard:2016tnd}
Jonah~E. Bernhard, J.~Scott Moreland, Steffen~A. Bass, Jia Liu, and Ulrich
  Heinz.
\newblock {Applying Bayesian parameter estimation to relativistic heavy-ion
  collisions: simultaneous characterization of the initial state and
  quark-gluon plasma medium}.
\newblock \emph{Phys. Rev.}, C94\penalty0 (2):\penalty0 024907, 2016.
\newblock \doi{10.1103/PhysRevC.94.024907}.

\end{thebibliography}
\end{document}